\documentclass[preprint,12pt]{elsarticle}
\usepackage{geometry}
\usepackage{subfig}
\usepackage{algorithm}
\usepackage{algorithmic}
\usepackage{amssymb}
\usepackage{amsmath}

\begin{document}

\begin{frontmatter}



\title{Dynamic Control Allocation between Onboard and Delayed Remote Control for Unmanned Aircraft System Detect-and-Avoid}


\author[a]{Asma  Tabassum\corref{author}}
\author[a]{He Bai}

\cortext[author] {Graduate Research Assistant \\\textit{E-mail address:} asma.tabassum@okstate.edu}
\address[a]{Department of Mechanical and Aerospace Engineering, Oklahoma State University}

\begin{abstract}
This paper develops and evaluates the performance of an allocation agent to be potentially integrated into the onboard Detect and Avoid (DAA) computer of an
Unmanned Aircraft System (UAS). We consider a UAS that can be fully controlled by the onboard DAA system and by a remote human pilot. With a communication channel prone to latency, we consider a mixed initiative interaction environment, where the control authority of the UAS is dynamically allocated by the allocation agent. In an encounter with a dynamic intruder, the probability of collision may increase in the absence of pilot commands in the presence of latency. Moreover, a delayed pilot command may not result in safe resolution of the current scenario and need to be improvised. We design an optimization algorithm to reduce collision risk and refine delayed pilot commands. Towards this end, a Markov Decision Process (MDP) and its solution are employed to create a wait time map. The map consists of estimated times that the UAS can wait for the remote pilot commands at each state. A command blending algorithm is designed to select an avoidance maneuver that prioritizes the pilot intention extracted from the pilot commands. The wait time map and the command blending algorithm are implemented and integrated into a closed-loop simulator. We conduct ten thousands fast-time Monte Carlo simulations and compare the performance of the integrated setup with a standalone DAA setup. The simulation results show that the allocation agent enables the UAS to wait without inducing any near mid air collision (NMAC) and severe loss of well clear (LoWC) while positively improve pilot involvement in the encounter resolution.

\end{abstract}


\end{frontmatter}

\baselineskip 18 truept

\section{Introduction}

\label{sec: intro}

In this era of cutting edge Artificial Intelligence, implementation of  intelligent software agents and advent of advanced algorithms are making autonomous vehicles more adaptive to uncertainties and anomalies. Human computer interaction studies aim to leverage a human's ability to aid in the execution of  autonomous actions while ensuring the best experience for a human operator. The balance between the two different paradigms, namely autonomous control and human control, falls into the mixed-initiative interaction paradigm~\cite{hearst1999mixed}. Marti et al.~\cite{hearst1999mixed} describe mixed-initiative interaction as a \textit{"Flexible interaction strategy in which each agent (human or autonomous system) contributes what it is best suited at the most appropriate time"}. This allows an autonomous system working with a human to develop a shared understanding of the goals and of contributing to the problem solving in the most appropriate way. In a generic situation, the disengage and reengage of the autonomous system can be initiated manually. However, for a safety critical system in the presence of anomaly, uncertainty and obstacles, manual transition of the command may lead to undesired consequences. Often times an operator requires some advisory and needs a considerable amount of time to take over control as the situation demands~\cite{mok2015emergency}, which may lead to non-smooth or bumpy transitions~\cite{sarter1997automation}. This establishes the necessity of designing efficient algorithms to allocate and blend the commands based on the quantitative as well as the qualitative analysis of the proposed actions. When designing such a controller scheme, the ultimate goal, constraints and cost functions vary in a wide range depending on the applications. 

For aerospace applications, optimization and control design for a standalone airborne system itself demands more scrutiny and requires guaranteed operability in a dynamic environment. This becomes more challenging for a UAS, where a remote pilot does not have the provision of analyzing the dynamic environment onboard. Therefore, the task to develop a control scheme that can dynamically allocate control authority between the pilot and the DAA system as well as adjust both commands to guarantee safe maneuvers and adaptation to uncertainties is challenging. 

In this paper, we consider a Detect-and-Avoid problem, where an ownship UAS is in an active encounter with a dynamic intruder. The UAS is remotely controlled by a pilot at the Ground Control Station (GCS). In the presence of the communication latency, the UAS needs to make decisions on 1) whether it can wait for pilot commands given the current encounter and 2) how to integrate the delayed pilot commands when received.
To answer these questions, we employ a dynamic control allocation agent to ensure effective authority transition in the encounter. Without redesigning the onboard DAA system, we implement a decision-theoretic approach to
effectively increase the remote pilot’s contribution within safe operational conditions and enhance human-machine user experience. In particular, we employ a Markov Decision Process (MDP) to generate a dynamic waiting strategy for the ownship UAS when pilot commands are unavailable due to latency. One of the benefits of waiting is that if the pilot's maneuver commands are received while the UAS is waiting, the intent of the maneuver, e.g., the maneuver types, can be assimilated into the onboard DAA system to generate a maneuver similar to the pilot commands.

We evaluate the competency of the MDP wait map using $10,000$ fast-time Monte Carlo simulations. An extensive data analysis is carried out on the simulation data to  quantify the performance and measure the risk associated with the allocation agent. Our results demonstrate noteworthy improvement in the pilot command reception without inducing any near-mid-air collision instances and enhancement of pilot involvement in decision making in the encounter resolution. Utilization of the pilot command has increased by $15.81\%$ with the integrated control allocation agent. On average, the implemented command blending algorithm enables execution of commands in the pilot preferred approach in $84\%$ of the instances. This paper generalizes the 2D MDP approach in~\cite{li2019markov} to 3D and extends~\cite{tabassum2020optimizing} by providing a large scale simulation study and analysis to validate the proposed approach.


The rest of the paper is structured as follows. The rest of the Section \ref{sec: intro} discusses  the literature and the contribution of this paper. Section \ref{sec: formulation} presents the problem definition, mathematical formulation of the MDP, and the command blending algorithm. Section \ref{sec: simulation} provides the simulation parameters, encounter designs and analysis metrics. The simulation results are presented and analyzed in Section~\ref{sec: results}. Conclusions are provided in Section \ref{sec: conclusion}.

\subsection{Literature Review} 
In a shared control architecture, both the autonomous system and human operator have full authority to steer a dynamical system. In the recent years, several shared control architectures have been proposed based on various coordination and interaction methods, including control authority switch, supervise-assist and command blend. To have an efficient coordination between human and autonomy, studies have been carried out on the transition of the control authority. These studies presented different approaches such as emergency stop, adaptive control, Support Vector Machines (SVM), and Model Predictive Control (MPC) to address different scenarios in diverse operating conditions. Control authority allocation between an autonomous system and a remote human operator demands dynamic optimization with guaranteed operability and safety under a wide range of operating conditions, constraints and uncertainty. Several studies~\cite{erlien2015shared,tran2016driver,walch2016towards,farjadian2016resilient,thomsen2019shared} address this problem depending on the system specifications by utilizing  existing methods with adaptations for ground and airborne vehicle. The authors~\cite{erlien2015shared} develop a simultaneous steering control algorithm using adaptive MPC such that the controller shares control with the driver in a minimally invasive manner while avoiding obstacles and preventing loss of control. The authors~\cite{tran2016driver} employ control switching between a driver and an autonomous system that is triggered by the driver drowsiness analysis. In~\cite{walch2016towards}, the authors propose to use an emergency stop as a fallback solution to maintain the vehicle’s safety in case the human operator never responds to a proposition from automation. The authors in~\cite{farjadian2016resilient} propose a switching control method for an autonomous flight controller where a human pilot takes over based on  the pilot's perception of the Capacity of Maneuver of the actuator. Later, they propose transitioning authority from the autopilot to a more advanced autopilot following the pilot's perception of an anomaly~\cite{thomsen2019shared}. 

Researchers have also considered human as supervisor and autonomous system as assistant, e.g.,~\cite{chipalkatty2011human,chipalkatty2013less,penizzotto2015human,shia2014semiautonomous,eraslan2020shared,thomsen2018shared,franchi2012shared,anderson2013intelligent},  In~\cite{chipalkatty2011human}, a theoretic controller is formulated for shared control of a quadruped rescue robot that composes human inputs with autonomous controls to guide human controlled leg positions. In a subsequent work~\cite{chipalkatty2013less}, the authors adopt a ``prediction-human” feedback loop by applying a dual-mode MPC technique and search-and-rescue inspired human operator trial. In~\cite{penizzotto2015human}, online evaluation of the human operator’s risk-based performance is employed for bilateral tele-operation of mobile robots to make compensation for their commands in the event of poor operator performance. To keep the safety of the vehicles, the authors~\cite{shia2014semiautonomous} propose to model the driver from empirical observations and incorporate the model into a control framework to correct the driver’s input. In~\cite{eraslan2020shared}, the authors propose a model based on supervisor authority over autonomous flight controller during anomaly and validate the model with human-in-the loop simulations. The authors in~\cite{thomsen2018shared} adopt a shared supervisory architecture in which the adaptive autopilot changes the controller structure and responds accordingly in the event pilot detects any anomaly in flight. In~\cite{franchi2012shared}, a control framework and its associated experimental test bed for the bilateral shared control of a group of QR UAVs are investigated where a topological motion controller increase the tele-presence of human assistants.

A promising shared control architecture is the command blend architecture~\cite{vanhooydonck2003shared,philips2007adaptive,goil2013using,desai2005blending,storms2014blending,inagaki2003adaptive} where the commands of the human operator and the autonomous system are fused to control the system. A wheelchair shared controller is presented in~\cite{vanhooydonck2003shared} where the navigation path was adapted with user intention. Reference~\cite{philips2007adaptive} presents a brain-computer interface for shared control of an wheel chair to perform goal directed navigation and assist human adaptively. The authors in~\cite{goil2013using} propose to use a Gaussian Mixture Model to learn weighing coefficients to blend the control. Input mixing control is adopted by some studies~\cite{desai2005blending,storms2014blending} where human-autonomous commands are blended to determine the final command. The authors in~\cite{desai2005blending} present a methodology of sliding scale autonomy to allow a user to adjust the autonomy of the flying robot. However the human has to know the sliding scale as the environment changes. In~\cite{storms2014blending}, the authors calculate a scale factor by minimizing a cost function that incorporates human command, obstacle potential field and obstacle avoidance controller.  The scalar blending factor  for human-autonomous command blending,  $\alpha$, is used to denote the degree of the authority. The value of $\alpha$ can be set $0$ or $1$ or changed adaptively~\cite{inagaki2003adaptive}. The percentage of $\alpha$ indicates how much control authority should be allocate to human-machine over different operating scenarios. Apart from shared control concept, fault tolerant adaptive controller~\cite{takase2020pilot} has been studied with a pilot in the loop simulation to reduce pilot workload during faulty situations. The authors in~\cite{bucolo2020bifurcation} consider Pilot Induced Oscillations (PIOs) from the point of view of nonlinear dynamics and nonlinear control in aircraft. 

As UAS applications continue to expand, researches have been carried out on the path prediction, tracking and control as well as collision avoidance algorithm development~\cite{kang2020model, cao2020discrete,wang2020three,zhang2019fast,pierpaoli2018uav,radmanesh2020towards, hamada2018receding, temizer2010collision}. Detect-and-Avoid concept of operation is illustrated in~\cite{do2017365} and several studies~\cite{tabassum2019probabilistic, weinert2018well, wu2020detect, wang2020three,lu2020simulations} specifically focus on the safety critical analysis of a standalone DAA with different methodologies as well as design of cognitive pilot-machine interface ~\cite{lim2018cognitive,lim2019cognitive}. When implementing a  shared architecture with UAS DAA, it is important to consider any inconsistency or latency that exists in the communication channel, since communication is crucial for sending surveillance information  to ground control station and receiving pilot command.  According to~\cite{chen2007human}, latency can be described as one of the most significant factors for instability and performance degradation. In \cite{palacios2013short}, the authors show a simulated communication loss at air traffic network increases conflict rates beyond the baseline level within 1 minute of the failure and that increased by at least a factor of 4 within 5 minutes of the communication failure. For flight control system, latency or delay influences flying quality and reliability of the operation and requires substantial effort to restore~\cite{ionita1996input}. Also in~\cite{kim2016effects}, the authors demonstrate how long range network latency causes reduced control performance as well as instability of the UAS. The authors design classical PID to compensate for the effects of time delay and stabilize the vehicle. A sensitivity analysis has been carried out in~\cite{laupre2019sensitivity} that analyzes the intolerances
of possible delays in control-input command with respect to the navigation performance on a fixed-wing
unmanned aerial system and states that a rapid growth of position error even for delays as small as ten of milliseconds may directly affect the situational awareness. To compensate for the latency in bilateral tele-operation of a UAS, the authors in~\cite{salinas2015complete} propose to use proportional plus damping injection (P+D) controller to increase the system stability. In~\cite{lam2007collision}, the authors consider obstacle avoidance for tele-operated UAVs with time delays and evaluates the effectiveness of using wave variables in haptic feedback to improve operators’ avoidance performance. The authors in~\cite{armah2017analysis} demonstrate how to use analytical solutions for ODEs and DDEs to
estimate the time delay in Internet-based feedback control and quadrotor types of
UAS. A feedback augmentation method is illustrated in~\cite{cox2019predictive} where the estimate of the change in the vehicle state due to the commands that are yet to affect the feedback is computed. This feedback received by the pilot is modified to reflect the predicted
change. This way the pilot becomes aware of the effect of the control inputs immediately.

\subsection{Contribution}

Although the existing control allocation and shared control architectures have considered diverse ranges of operating conditions, including vehicle dynamics, anomalies and sensor uncertainties, communication latency has not been considered extensively, which is the main focus of this paper. The existing literature for airborne platforms consider the supervisor-assist~\cite{eraslan2019shared,thomsen2018shared,  farjadian2017bumpless} control architecture in the presence of anomalies where we consider a allocation-blend architecture that connects degraded pilot's situational awareness unlike the standalone baseline setup. The proposed architecture is capable of both control transition and command blend in a dynamic environment with communication latency. The control allocation agent dynamically allocates the control authority between the onboard DAA and the pilot and is separate from the autonomous system and the human operator. Although a control allocation agent has been successfully implemented in many complex distributed control plants including but not limited to electro-hydraulic driven devices and industrial robots~\cite{chen2012robust, zhou2016predictive, vermillion2010predictive, zhang2020dynamic}, airborne vehicles~\cite{guo2020finite, lang2020control, zhao2019fast}  and in multi-agent network~\cite{buzorgnia2018follower} implementation in an unmanned aircraft mixed initiative application is very limited. This paper defines a mixed initiative infrastructure that utilizes MDP wait maps, current relative states, pilot commands and DAA commands to safely authorize pilot or DAA to take over the control as well as incorporate the pilot intention. The developed wait map is designed to ensure safe navigation when the allocation agent holds the authority and blend pilot intended maneuver into the resolution. The efficiency of the MDP wait map and competence of the allocation agent are investigated through $10,000$ fast-time Monte Carlo simulations that encompass a wide range of encounter geometry. The results show positive improvement in pilot command involvement and produce an average increase of $15.81\%$ of the utilization of pilot commands  than a standalone DAA architecture. Furthermore, none of the simulations has induced  NMAC instances.

\section{Problem Formulation}\label{sec: formulation}
Consider an ownship UAS whose communication channel to and from the GCS suffers from latency during an encounter with a dynamic intruder aircraft. The round-trip latency duration can be in the range of 0.2 to 10 seconds \cite{do2017365}. As the onboard surveillance system detects the intruder, the surveillance information   is  transmitted to the pilot at the GCS. The remote pilot from the GCS takes a maneuver decision and sends commands back to the UAS. We assume that the onboard DAA system is also capable of generating and executing avoidance maneuvers. However, these maneuvers may be significantly different from the maneuvers generated by the pilot. On the other hand, the pilot command may not be received by the UAS prior to a resolution maneuver time frame. To improve pilot-UAS interaction experience, it may be desirable to hold control authority for pilot over computer. However, in the event of latency, waiting for remote pilot's commands may significantly increase the risk of collision. 

We propose a control allocation agent that utilizes a prior waiting map solved from a MDP model. The MDP is developed in 3-D dimension and based on the ownship and intruder initial state and the intruder dynamic motion model. The solution to the MDP acts as prior information for the agent to decide how long the UAS can wait for the pilot's commands before the onboard DAA is authorized to take over. We next discuss how the MDP is formulated and the design of the control allocation agent. 
\subsection{Formulation of the MDP}\label{sec:state space}
A generic MDP can be defined as a tuple $(S,A,P,R)$ \cite{puterman2014markov}, where $S$ is the state space, $A$ is the action space, $P$ is the transition probability and $R$ is the reward function. We describe the relationship between the UAS and an intruder in a relative coordinate system. A kinematic model for both the UAS and the intruder is adopted from \cite{beard2012small}. In our study, the state space has six states: relative distance in the \textit{x} dimension, $\Delta X$, relative distance in the \textit{y} dimension, $\Delta Y$, relative altitude, $d_{h}$, intruder horizontal speed, $v_{i}$,  relative vertical velocity, $V_{h}$, and intruder heading, $\theta_{i}$. 
The states are updated as
\begin{align}
 \Delta X (t+ \Delta T)=\Delta X (t)+V_{rx}(t) \Delta T  \label{eq:x}  \\
  \Delta Y (t+ \Delta T)= \Delta Y (t)+V_{ry}(t) \Delta T \\ \label{eq:y}
  d_{h}(t+\Delta T)=d_{h}(t)+ V_{h}(t) \Delta T\\ \label{eq:z}
  v_{i}(t+\Delta T)= v_{i}(t)+\Delta v_{i}(t)\Delta T \\ \label{eq:intru vel}
   \theta_i(t+\Delta T)=\theta_{i}(t)+ \Delta \theta_{i}(t) \Delta T \\ \
    \ V_{h}(t+ \Delta T)=V_{h}(t)+ \Delta v_{h} (t) \Delta T, \label{eq:vertical rate}
 \end{align}
where $\Delta T$ is the transition time and we set $\Delta T= 1 $ second.
The values of  $V_{rx}$ and $V_{ry}$ are calculated as
\begin{align}
     V_{rx} (t) &=v_{i}(t)\cos\theta_i(t)-v_{o} (t) \cos\theta_o (t) \\
      V_{ry}(t) &=v_{i}(t) \sin\theta_i (t) -v_{o} (t) \sin\theta_o (t)
\end{align}where $v_{o}$ and $\theta_{o}$ are the ownship horizontal speed and heading, respectively.  

State transition probability depends on the action taken by the UAS and the probability of the intruder dynamics. We use a preset intruder motion model to compute the transition of the MDP. Table \ref{table 1} illustrates the intruder motion model, $m_{i}$. Equation~\ref{eq:x}--\ref{eq:vertical rate} can be described by a generic discretized state space model as:
\begin{equation} \label{eq:state space}
    S(k+1)=f(S(k),u_{a}(k), \Delta v_{i}, \Delta\theta_{i}, \Delta v_{h})
\end{equation}
where $S(k)\in R^{6\times 1}$ is the current state, $u_{a}(k)$ is the current input command that includes the current ownship airspeed, heading and vertical velocity, which are assumed available, and $\Delta v_{i}, \Delta\theta_{i}, \Delta v_{z}$ are drawn from the intruder motion model $m_{i}$ in  Table \ref{table 1}. Equation~\ref{eq:state space} describes the discrete state transition depending on the current state $S(k)$ given $u_{a}(k)$ and the intruder maneuvers randomly drawn from Table~\ref{table 1}. 
 
 \begin{table*}[!t]
\renewcommand{\arraystretch}{1.3}
 	\caption{Intruder Motion Model, $m_{i}$} \label{table 1} 
 \centering
\tiny
 \begin{tabular}{c c c c c c}

 Vertical dynamics& Probability &Horizontal dynamics & Probability & Turn rate dynamics&  Probability  \\
 	 \hline\hline 
 $\Delta v_{h}$ (m/s$^2$) & $P(\Delta v_{h})$ &  $\Delta v_{i}$ (m/s$^2$) & $P(\Delta v_{i})$ & $\Delta\theta_{i}$ (deg/s) & $P(\Delta \theta_i)$ \\ [0.3ex]

 	-5 & 0.15 & 	-10 & 0.10	& -5 & 0.20\\
 	-3 & 0.20 & 	-5 & 0.15  & -2.5 & 0.20\\
 	0 & 0.30 &  	-2.5 & 0.15  &  	0 & 0.20\\
 	3 & 0.20 &  	0 & 0.20 &  	2.5 & 0.20\\ 
 	5 & 0.15 &  	2.5 & 0.15 &  	5 & 0.20\\
 & &              	5 & 0.15 & & \\
& &                	10 & 0.10 & &\\
 	
 \end{tabular}
 \end{table*}
We include two more special states named \textit{Out} state and the \textit{LoWC} state. The \textit{Out} state is the state after the UAS takes an evasive maneuver and after this the MDP will be terminated. The \textit{LoWC} state is the state where the UAS and the intruder have a distance smaller than a preset threshold. It is an event in which a UAS is in close proximity with another aircraft such that the following three conditions are concurrently true \cite{do2017365}:
\begin{align}
d_{h}\leq d^*_h \\
HMD \leq HMD^*\\
0 \leq \tau_{mod} \leq \tau^*_{mod}
\end{align} where the asterisked parameters are thresholds later described as $LoWC_{threshold}$ and non-asterisked parameters are measured values, $d_{h}$ is the vertical separation, HMD is the horizontal miss distance and $\tau_{mod}$ is the modified $tau$. $HMD$ is the projected separation in the horizontal dimension at the predicted closest point of approach. The mathematical definition is provided in~\cite{do2017365}.

The action space $A$ consists of  two decisions: evasive maneuver and wait. We define $A=\{0,1\}$, where $A=0$ refers to wait and $A=1$ refers to taking an evasive maneuver from the onboard DAA system. The MDP compares state transitions towards the Out state with other discrete states in terms of reward and decides if an evasive maneuver should be taken. For simplicity, we assume that the intruder will take only the vertical or the horizontal maneuver at one time. For the horizontal maneuvers, we consider each possible pair of horizontal speed and turn rate that may be taken by the intruder and compute their probabilities as the product of their probabilities in the intruder dynamic model, $m_{i}$ in Table \ref{table 1}. Let $p_{i}$ be the probability of the intruder motion expressed as
 \begin{equation}
    p_{i} (m_{i})=\left\{\begin{array}{cl}
    \ P(\Delta v_{i})P(\Delta \theta_{i})  &\mbox{during horizontal maneuver,}\\
    \ P(\Delta v_{z}) &\mbox{ during vertical maneuver.}
    \end{array}\right.  
 \end{equation}
The transition probability for the actions are computed by:
 \begin{equation}
 P(S\sp{\prime}|S,A)=\left\{\begin{array}{cl}
    \sum_{m_{i}} P(S\sp{\prime}|S,A,m_i) p_{i} (m_{i})  &\mbox{if $A=0$}\\
    P(S\sp{\prime}=out|S,A,m_i)=1 &\mbox{if $A=1$}
    \end{array}\right.  
 \end{equation} where $m_{i}$ is the intruder motion model in Table~\ref{table 1}.  

 Each state transition is associated with a reward value depending on the action. The reward function for the wait action is based on the loss of well clear and given by
 \begin{equation}\label{eq:r_w}
 R_{w}(S\sp{\prime},S,A=0)=-\min(\frac{C_{1}}{HMD(S\sp{\prime}) },\frac{C_{2}}{d_{h}(S\sp{\prime})})+C_{3}\Delta T
 \end{equation}
where $C_1$, $C_2$, and $C_3$ are positive user defined parameters. In this study, we use $C_{1}$= $HMD^*$ = $1220$, $C_{2}$ =  $d_{h}^*$ =  $122$ and $C_{3}$ = 1. The first term in~\eqref{eq:r_w} penalizes close encounters and while the second term encourages waiting. The values of $C_1$, $C_2$ and $C_3$ are chosen such that whenever the encounter gets to too close or potentially violates well clear, the reward will be negative for the waiting action. At all times, taking evasive action will receive zero reward, i.e.,
\begin{equation}
 R_{e}(S\sp{\prime}, S,A=1)=0.
  \end{equation}
  Thus, at each time, the reward is given by 
\begin{equation}
    R (S\sp{\prime}, S,A) =\left\{\begin{array}{cl}
    R_{w}  &\mbox{if $A=0$}\\
    R_{e} &\mbox{if $A=1$.}
    \end{array}\right.
\end{equation}

The ownship UAS will receive a positive reward until the distance falls below a certain threshold. In the event when the intruder is in the close proximity, the reward for \textbf{$A=0$ } will be negative whereas the reward for \textbf{$A=1$} will be zero. Therefore, the UAS prefers an evasive action. 

To solve for the optimal action at each state, a value function is computed using dynamic programming based on value iteration. The value function is given by
\begin{equation}
    V^*(S(k))=\max_{a(k) \in A} \left(R(a(k))+\gamma \sum_{S(k+1)\in S}  P(S(k+1)|S(k),a(k))  V^*(S(k+1)) \right) \quad
\end{equation} $\forall S(k)\in S $, where $\gamma<1$ is the discounted factor and $R$ is the reward function. Using the value iteration algorithm, a discounted MDP is solved and the optimal action (wait or maneuver) is obtained at each state by maximizing the accumulated reward.
 
The wait time at each state is calculated using weighted transition probabilities. For every state, we propagate every possible path to the $LoWC$ state and obtain the corresponding LoWC probability. Assuming that the path is acyclic, the expected waiting time for that state is calculated as the average of the waiting times of all the paths leading to LoWC weighted by their probabilities. A wait map is created by computing the wait time at each state in the map.

\subsection{Design of the Control Allocation Agent} \label{sec:blend}

\begin{figure}[!h]
    \centering
    \includegraphics[width=.5\textwidth]{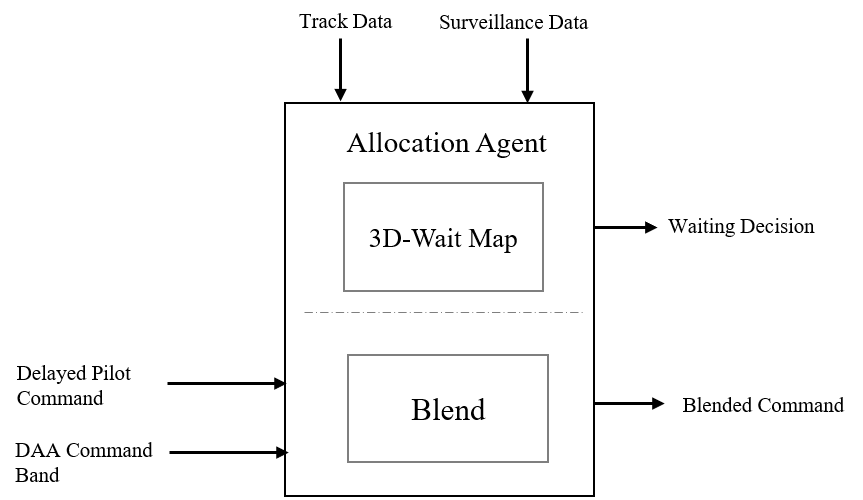}
    \caption{Input/Output diagram for the allocation agent.}
    \label{fig:agent block}
\end{figure}

We employ an allocation agent which functions separately from the human or the DAA and works as a high level decision maker. 
The input/output information of the allocation agent is presented in Figure \ref{fig:agent block}. 
The waiting time map  developed from the MDP solutions acts as prior information for the agent. The command allocation at any time depends on the status of pilot command reception, projected loss of well clear $ProjL(.)$ within a look ahead time of $N$ seconds, a safety metric of command $SM(.)$,  estimated maneuver time and optimal wait time at the current state. The projected loss of well clear is a function that predicts a loss of well clear within the next $N$ seconds and expressed as
\begin{equation}
    ProjL(i) = \left\{\begin{array}{cl}
    \ 1   &\mbox{ if $ \{HMD(i), d_{h} (i) , \tau_{mod}(i)\} \leq LoWC_{threshold}  $}\\
    \ 0 &\mbox{ if $\{HMD(i), d_{h} (i) , \tau_{mod}(i)\} > LoWC_{threshold}  $} ; \quad i=1,\cdots, N.
    \end{array}\right.  
\end{equation}

When a pilot command is not available, the agent either decides to wait for the pilot command and maintain course, i.e., trajectory from previous pilot commands, or allocate the authority to the onboard DAA for conflict resolution. The wait and no wait decisions are made as 
\begin{equation}\label{eqn: waiting_stretagy}
    W(k) =\left\{\begin{array}{cl}
    Wait  &\mbox{if $w(S(k)) \geq W_{T})$ \mbox{ and }  $ProjL(i)=0,~\forall i=1,\cdots,N$}\\
    No~ Wait &\mbox{if $w(S(k)) < W_{T})$ \mbox{ or }  $ProjL(i)=1$,  $\exists i\in\{1,\cdots,N\}$}
    \end{array}\right.
\end{equation}
where $w(S(k))$ is the optimal wait time at state $S(k)$ retrieved from the wait map, and $W_{T}$ is a predetermined wait time threshold. Such a threshold can be chosen based on an estimate of communication latency from available historical data or from a model. Latency of a specific channel can be mathematically modelled \cite{nguyen2011modeling,intharawijitr2016analysis,yang2020multi} or predicted using statistical analysis. Also, the threshold can be determined from available data of the specific aircraft transmitter-receiver. In our simulations, we have chosen $W_{T}=4$ sec given delays between 0.2 and 10 seconds.

Prioritizing safety, Equation~\ref{eqn: waiting_stretagy} presents two logical decisions on waiting or not. If at any time, the wait time at the current state is greater than or equal to time threshold and no loss of well clear is projected within the next $N$ seconds, the agent holds the authority for pilot. Although the wait time map was created with $LoWC$ penalty, we have added the $LoWC$ projection logic to create safeguard. On the other hand, if at any time, the wait time is less than the threshold time which implies a delayed pilot command may not be available within a specific wait time, or there is a projected $LoWC$ within $N$ seconds, the agent authorizes the onboard DAA to take over the 
control.


When the UAS receives delayed pilot command, the agent optimizes which command to execute  to maintain safety as well as enhanced piloting experience. If the UAS receives a delayed pilot command and determines that it is safe to execute that command, the delayed pilot command is implemented. In the case that the received pilot command is not safe evaluated by the safety metric SM, we adopt a blending approach where the preferred maneuver by the pilot is assimilated into DAA maneuver selection. Thanks to the waiting action, the pilot preference can be extracted from the received (delayed) command. Let the received pilot command and the DAA command be $u_{p}$ and $u_{DAA}$, respectively. The command allocation algorithm is presented in Algorithm \ref{alg:1}.

We now discuss how delayed pilot maneuver commands are employed when they are deemed unsafe. Preference of the pilot maneuver is stored as paired information $(ax,d)$, where $ax$ refers to the axis or plane of the maneuver, i.e., horizontal or vertical maneuver, and $d$ refers to the direction of the maneuver, i.e., left or right in case of horizontal maneuvers and up and down in case of vertical maneuvers. The agent first searches for a safe avoidance maneuver that matches the pilot preference in terms of maneuver axis and direction. The found maneuver may be more aggressive than the received pilot command due to the latency. If no such safe maneuver is found in the intended direction, the UAS then searches for a safe avoidance maneuver that matches the intended axis. When no safe maneuver can be found to match the intention, the UAS will take any safe maneuver produced by the onboard DAA system. The blending algorithm is described in Algorithm \ref{alg:2}.




\begin{algorithm}[h!]
	\caption{Command allocation algorithm }\label{alg:1}
	\textbf{Input}: command reception flag, pilot command $u_{p}$, DAA commands $u_{DAA}$, current intruder state information, current ownship state information.\\
	\textbf{Output}: Avoidance maneuver for the ownship, $u_{a}$ \\
\textbf{if} pilot command not received

\hspace{.25 in} \textbf{if} $W(k)== Wait$

\hspace{.5 in} $u_a \gets u_p$ 

\hspace{.25 in} \textbf{else if } $W(k)==No~Wait$

\hspace{.5 in}  $u_a \gets u_{DAA}$

\textbf{else} pilot command received 

\hspace{.25 in} \textbf{if} $SM(u_{p})$ $>$ 0

\hspace{.5 in}   $u_a \gets u_p$

\hspace{.25 in } \textbf{else if } $SM(u_{p})<0$

\hspace{.5 in} invoke \textbf{Algorithm \ref{alg:2}}

\end{algorithm}


\begin{algorithm}[h!]
	\caption{Command blending algorithm}\label{alg:2}
	\textbf{Input}: Extracted $(ax,d)$, current ownship and intruder state information\\
	\textbf{Output}: Avoidance maneuver for the ownship, $u_{a}$
	\begin{enumerate}
\item Based on the current ownship and intruder state information, search for a safe avoidance maneuver from the onboard DAA system that matches $(ax,d)$
\item \textbf{if} success \textbf{then} \textit{return} the found maneuver as $u_{a}$,
\item Based on the current ownship and intruder state information, search for a safe avoidance maneuver from the onboard DAA system that matches $ax$ only
\item \textbf{if} success \textbf{then} \textit{return} the found maneuver as $u_{a}$,
\item Based on the current ownship and intruder state information, \textit{return} any safe avoidance maneuver as $u_{a}$.
	\end{enumerate}
\end{algorithm}	

\section{Simulation Parameters and Analysis Metric}\label{sec: simulation}

In the development of wait maps, all distances are considered in the relative coordinate system. The UAS position is $(0,0,0)$ m and has a constant airspeed of $55$ m/s with a heading of $0$ degree. The state space ranges from $-1500$ m to $1500 $m for $\Delta X$ and $\Delta Y$, $-200$ m to $200 $ m for $d_{h}$,  $70$ m/s to $300 $m/s for  $v_{i}$ and $-5$ m/s to $ 5 $ m/s for $V_{h}$.  The numbers of bins for discretizing the states are $12,~12,~4,~5,~5,~5$, respectively. For any transition, whenever the HMD is less than the HMD threshold and the current vertical separation is less than the vertical separation threshold, the reward will be negative for the waiting action and an evasive action will be taken. We use a publicly available toolbox \cite{toolbox} to solve the MDP. After solving the MDP, we have the optimal decision at each state and then calculate the corresponding wait time.

\subsection{Generate the Wait Maps}
\begin{figure}[h]
     \centering
     \includegraphics[width=.45\textwidth]{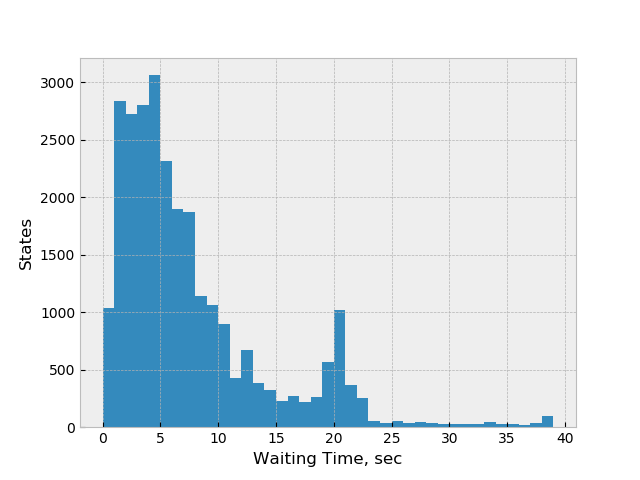}
     \caption{Histogram: Waiting time through out states}
     \label{fig:hist_wait}
 \end{figure}

\begin{figure}[h!]
 
     \centering
     \includegraphics[width=.45\textwidth]{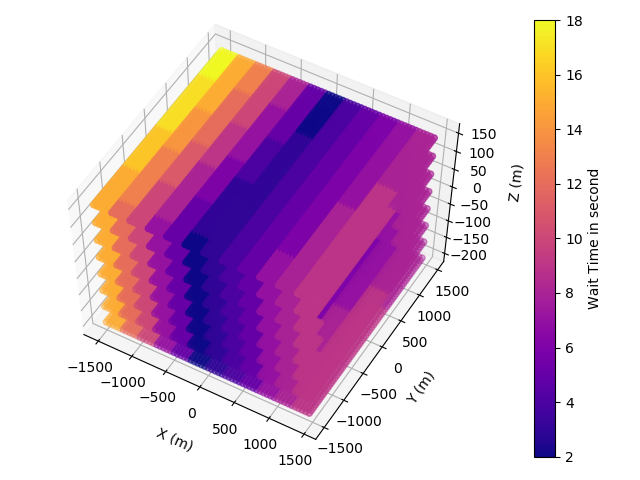}
     \caption{An example of 3D waiting map}
     \label{fig:3d map}
 \end{figure}

By changing the intruder initial conditions, we obtain different wait maps. Figure \ref{fig:hist_wait} presents a histogram of wait time. The histogram mirrors a left skewed Gaussian distribution.  The maximum wait time is 40 seconds and the minimum wait time is 1 second. In most of the states the wait time time ranges between $5$ and $10$ seconds. A 3-D representation of a wait map is provided in Figure~\ref{fig:3d map}. 

The positive relative $X$ range depicts the intruder approaching the UAS, i.e., head-to-head encounters, and the negative relative $X$ range depicts overtaking encounters, i.e., the intruder is approaching the UAS from behind. The wait time is lower in head-to-head encounters than in overtaking encounters. This is because the dynamic collision volume is approaching the intruder and the closure rate is higher. The decrease in the wait time is observed as the relative $X$ distance reduces.

\begin{figure*}[!t]
\centering
\subfloat(a){\includegraphics[width=2.1in]{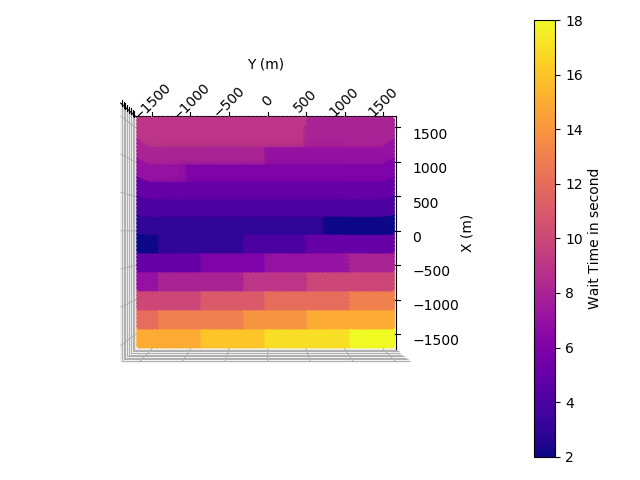}
}
\hfil
\subfloat(b){\includegraphics[width=2.1in]{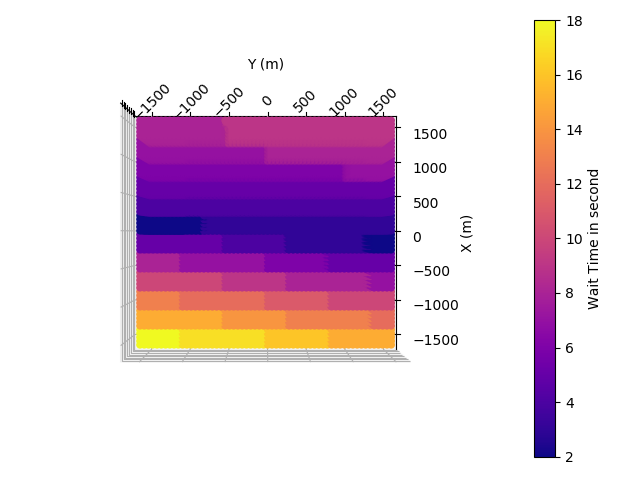}

}

\centering
\caption{Top View of a wait map,  (a) $v_{i}=155 $ m/s , $ \Delta \theta_{i}= -5 $ deg/s, (b) $v_{i}=155 $ m/s, $\Delta \theta_{i}= +5 $ deg/s. The wait map mirrors as the sign of the turn rate changes to the opposite.  }
\label{fig:top_view 1}

\end{figure*}

\begin{figure*}[!t]

\centering

\subfloat(a){\includegraphics[width=2.1in]{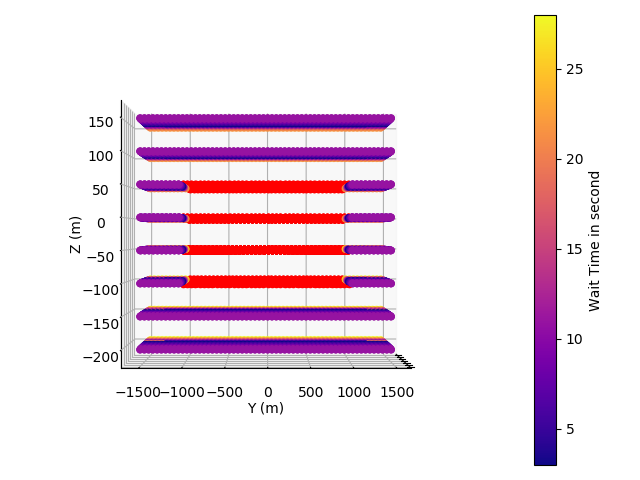}}
\hfil
\subfloat(b){\includegraphics[width=2.1in]{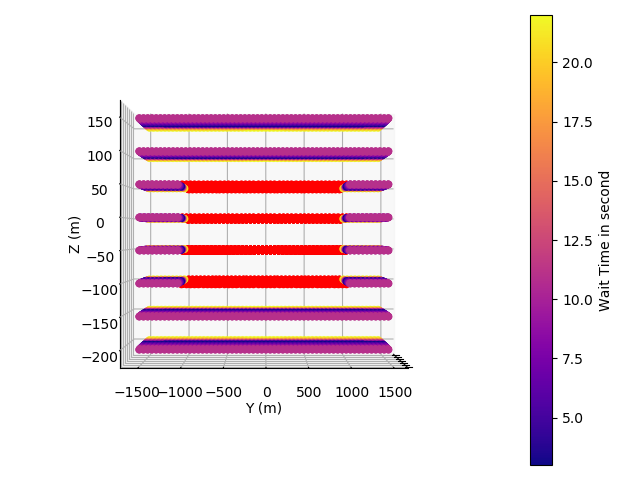}}

\caption{Front view of the wait maps for (a) $v_{h}= 3 $ m/s and (b) $v_{h}= 5 $ m/s within $\Delta X={[-400, -1500]}$.  The $\Delta X$ range has been chosen to illustrate the LoWC region (red area). An increase in the vertical velocity changes the waiting time.}
\label{fig: front view}
\end{figure*}

\begin{figure*}[!t]
\centering
\subfloat(a){\includegraphics[width=2.in]{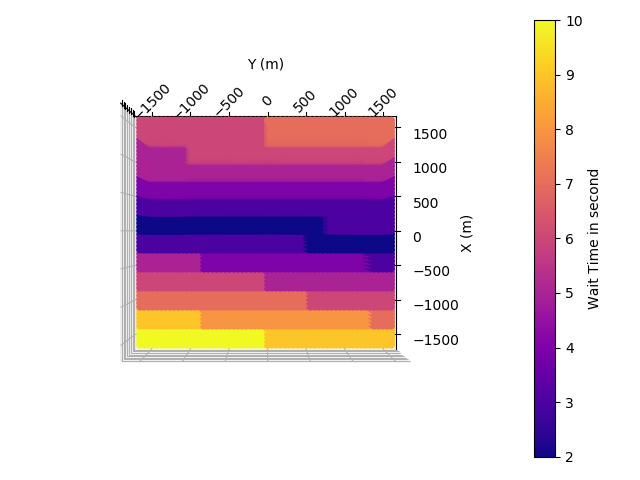}
}
\hfil
\subfloat(b){\includegraphics[width=2.1in]{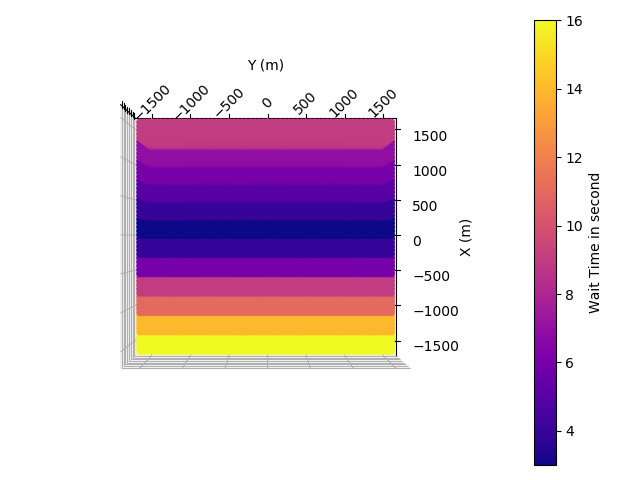}
}

\centering
\caption{ Top View of a wait map (a) $v_{i}= 300$ m/s , $\Delta \theta_{i}= 5 $ deg/s , (b) $v_{i}=155$  m/s , $\Delta \theta_{i}= 0 $ deg/s }
\label{fig:top_view 2}
\end{figure*}

Figure~\ref{fig:top_view 1} represents a top view of the wait map with a non-zero turn rate and illustrates how turn rate impacts the wait time. Comparing the two plots in Figure \ref{fig:top_view 1}, we see  how alternating turn angle changes the wait time, and the maps are mirror to each other. The positive initial turn leads to less wait time for intruders
appearing from the right in head-to-head encounters and from the left in overtaking encounters. On the other hand, a negative turn rate results in less wait time when the intruder converges from the left in  head-to-head encounters and from the right in overtaking encounters. The effect of change in vertical speed is presented in two different plots in Figure \ref{fig: front view} . The wait time also changes
as the relative horizontal speed changes. This can be seen from the two plots in Figure \ref{fig:top_view 2}, where the first plot has a similar condition to the second plot in Figure \ref{fig:top_view 1} except the change in the horizontal speed. The wait map in Figure~\ref{fig:top_view 1}-b has a higher wait time than the map in Figure \ref{fig:top_view 2}-a, since the horizontal speed in the latter plot is higher. Figure \ref{fig:top_view 2}-b represents a zero turn rate which results in a symmetric map. In the case  of a non-turning intruder, the wait time also increases with the horizontal speed. 

\subsection{Encounter Scenarios}
The encounter scenarios are representative scenarios from~\cite{do2017365}. We have generated $10,000$ pairs of aircraft trajectories that encompass a wide range of encounter geometry. For the generation of the trajectories, we consider the ownship flying straight north with a constant velocity. We define the incident angle $ia$ as the angle between the longitudinal axis of the ownship and the intruder in the counter-clockwise direction and the horizontal miss angle $hma$ as the angle between the longitudinal axis of the ownship and the position the intruder. To create turning geometries, we define another parameter `advance rate' that indicates the time to start turn. For example, an advance rate of $0.55$ indicates that a turning maneuver of the intruder happens at time $t=0.55T$ sec, where $T$ is the total encounter time. {We set $T=240$ sec.} The intruder parameters used to generate the waypoints are given in Table \ref{tab:encounter parameters}.

\begin{table}[h!]

   \caption{Intruder Parameters for Trajectory Generation }
    \label{tab:encounter parameters}
    \centering
    \begin{tabular}{c c c c}
    
         Parameters & Min & Max & Distribution  \\
         \hline \hline 
         Speed (m/s) & 70 & 400 & Gaussian(100,30) \\
         vertical velocity (m/s) & -10 & 10 & Gaussian(0,10) \\
         ia (deg) & 0 & 180 & 30\% 180, rest uniform \\
         hma (deg) & -110 & 110 & Gaussian(0,110)\\
         HMD (m) & 0 & 2750 & uniform\\
         VMD (m) & -915 & 915 & uniform \\
         turn rate (dps) & -5 & 5 & uniform \\         
         advance rate & 0 & 0.8 & Gaussian(0.5,0.5)
    \end{tabular}
\end{table}


\subsection{Implementation of the Wait Maps}
The wait time maps that we produce using the MDP are applied to a discrete event simulator. A baseline DAA algorithm with collision avoidance capability and the DAIDALUS~\cite{munoz2015daidalus} are integrated in the simulator. An unmanned aircraft pilot model~\cite{guendel2017model} is integrated in the simulator to mimic human-pilot response. Reference~\cite{guendel2017model} models the variability of human decision based on three different experimental setups. The model captures pilot response from the beginning of an encounter till safe avoidance. The pilot model uses DAIDALUS to choose avoidance maneuvers. The DAA algorithm onboard the ownship UAS has two primary functions: collision avoidance (CA), and DAA-Pilot mode of operation from the simulator. The combined operation mode allows DAA to override the pilot command in unusual events. We integrate the allocation agent with the onboard DAA. The allocation agent has the wait map information as resources and utilizes the maps during encounter in the events of communication latency. The agent is activated by default when the operation mode is DAA-Pilot with the provision of manually turning it off. The command blending algorithm in the DAA computer blends the delayed pilot commands with the onboard DAA commands. During the communication latency, the agent matches the current state with a state in the wait map and makes a decision according to~\eqref{eqn: waiting_stretagy}. The system waits for the optimal wait time extracted from the map until the control is being authorized to the onboard DAA or a delayed pilot command is received. The optimal time is updated when the state is transitioned. The second stage decision making starts when a delayed pilot command is received. The command blending algorithm discussed in Section \ref{sec:blend} is invoked to determine the maneuver options. We consider two different setup:
\begin{itemize}
    \item Baseline setup (B): DAA-Pilot mode 
    \item Integrated setup (I): Integrated Agent-DAA-Pilot mode.
\end{itemize}

\begin{figure*}[!t]

\centering
\subfloat(a){\includegraphics[width=2in]{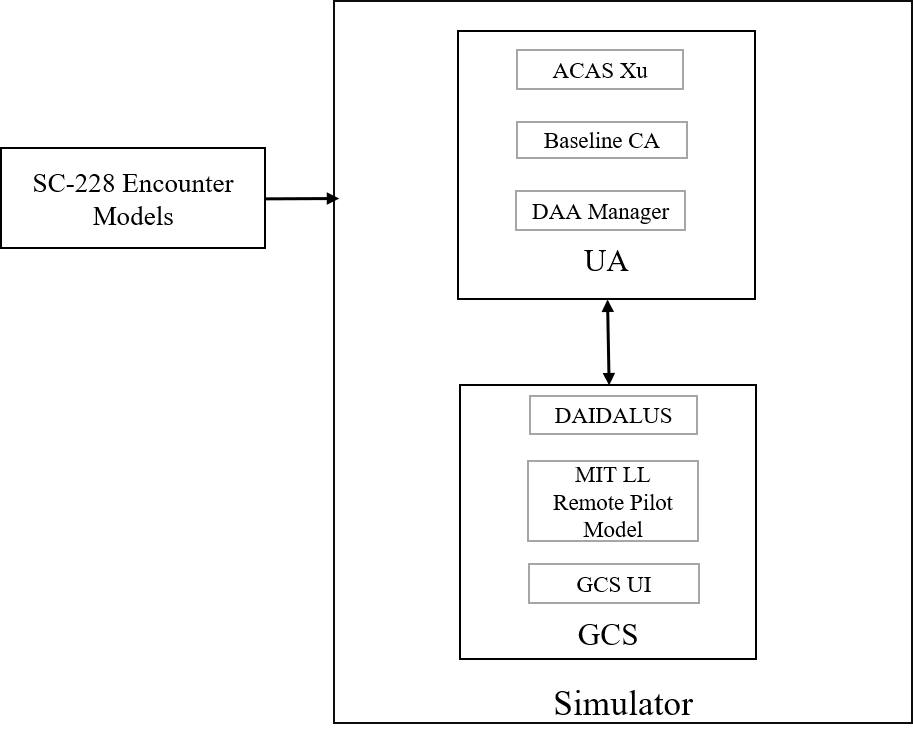}}
\hfil
\subfloat(b){\includegraphics[width=2.05in]{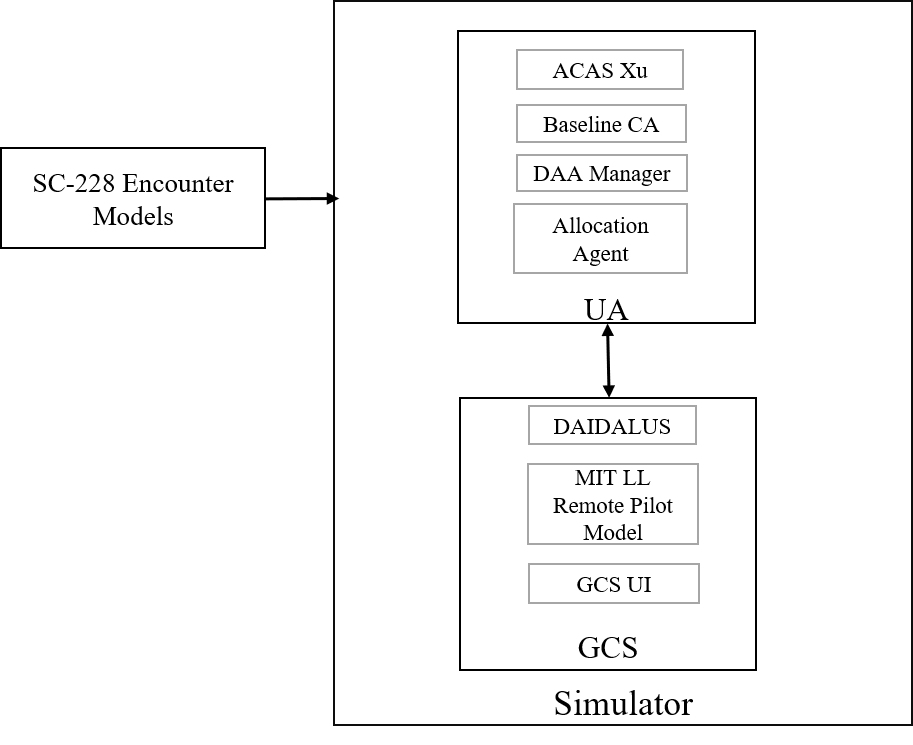}}

\caption{Block diagram of Simulator (a) Baseline Setup and (b) Integrated Setup with Agent}
\label{fig: Big block}
\end{figure*}

The Baseline setup is a standalone DAA setup as presented in Figure \ref{fig: Big block}-a and an integrated setup is the augmented allocation-blend architecture as presented in Figure \ref{fig: Big block}-b. We then design five groups of experiments with different configurations by varying the latency and restricting the use of wait maps within our two setups, as presented in Table \ref{tab:experiment sets}. The  latency is varied according to a Gaussian distribution with a mean around $5$ sec {and a standard deviation of 3 sec}. 

\begin{table}[h!]
   \caption{Experimental Groups.}
    \label{tab:experiment sets}
    \centering
    \begin{tabular}{c c c c c}
    
         Setups &  Groups & Latency type & Latency  (sec) & Wait time  (sec)  \\
         \hline \hline 
        Baseline & B-1 & Constant & 4 &  0 \\
        Baseline & B-2 & Varying & Gaussian & 0 \\
        Integrated & IC-1 & Constant& 5 & 5 \\
        Integrated & ID-1 & Constant & 4 & Map \\
        Integrated & ID-2 & Varying & Gaussian & Map\\
    \end{tabular}\\
    \footnotesize{*IC refers to the Integrated setup with a constant wait time and ID refers to the integrated setup with the wait maps.}
\end{table}

\subsection{Risk Analysis Metric}
To evaluate the risk associated with each configuration, we track the number of $LoWC$ and $NMAC$.  An NMAC occurs if 
\begin{itemize}
    \item $R\leq 500 $ ft,and
    \item $H\leq 120 $ ft,
\end{itemize}
where $R$ and $H$ are the horizontal and vertical separation. This pair of separation values is indicated as $NMAC_{threshold}$ later in the paper.
We quantify the risk as probabilities of $LoWC$ and probability of NMAC with Monte Carlo approximation. Define $P(LoWC)$ and $P(NMAC)$ as
\begin{equation}
    P(LoWC) \approx \frac{1}{K} \sum_{i=1}^K (\{HMD(S(i)),d_{h}(S(i)), \tau_{mod}(S(i))\} \leq LoWC_{threshold})
\end{equation}
and
\begin{equation}
    P(NMAC) \approx \frac{1}{K} \sum_{i=1}^K (\{ R(S(i)),H(S(i))\} \leq NMAC_{threshold})
\end{equation}
where $K$ is the number of sample trajectory points and $S(i)$ is the state vector at the $i$th time step. 

To examine the severity of LoWC, we utilize the metric `Penetration Integral' (PI)~\cite{do2017365}, which is a measure of penetration into the well clear volume and calculated as
\begin{equation}
    PI= \sum_{i=1}^K \min \left(\frac{4000-HMD(S(i))}{4000}, \frac{450-VMD(S(i))}{450}\right) \frac{(35-\tau_{mod}(S(i))}{35} \Delta T.
\end{equation}  PI is a scalar measure that quantifies the severity of each LoWC and distinguishes between prolonged severe LoWC and momentary LoWC. A LoWC value less than 2 are considered benign and momentary while a value greater than 10 is considered a significant LoWC~\cite{do2017365}. 

\section{Result Analysis }\label{sec: results}
We analyze the results of $10,00$ Monte Carlo simulations in this section. Our goal is to efficiently incorporate pilot command without jeopardizing safety. We investigate whether waiting positively improves pilot command incorporation into the system and if the prior wait maps help allocate the control authority safely. As the UAS waits for pilot commands, it continues to maintain the course, which triggers a question that how safe it is to maintain course in the presence of a dynamic intruder. We analyze the safety from the simulation results in the subsequent sections. We found our augmented allocation-blend architecture improves resolution of diverse encounter scenarios. For example, waiting retain controls for pilot and pilot successfully resolves the encounter. Blending command provides a better avoidance as well as recovery trajectory. Demonstration examples can be found in the supplementary material.

\subsection{Pilot Command Incorporation}
\begin{figure*}[!t]
\centering
\subfloat(a){\includegraphics[width=2in]{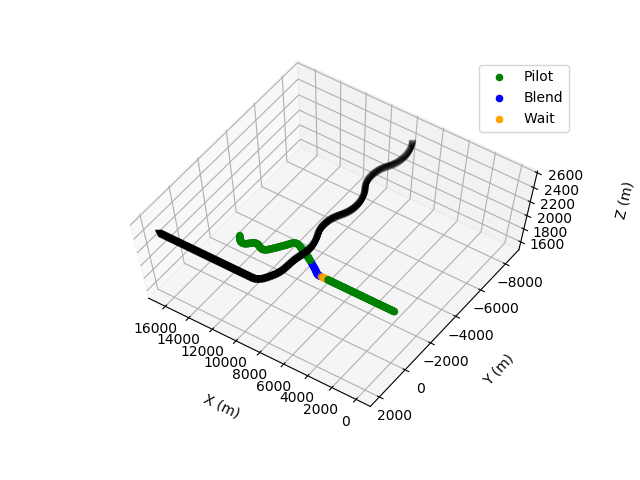}
}
\hfil
\subfloat(b){\includegraphics[width=2in]{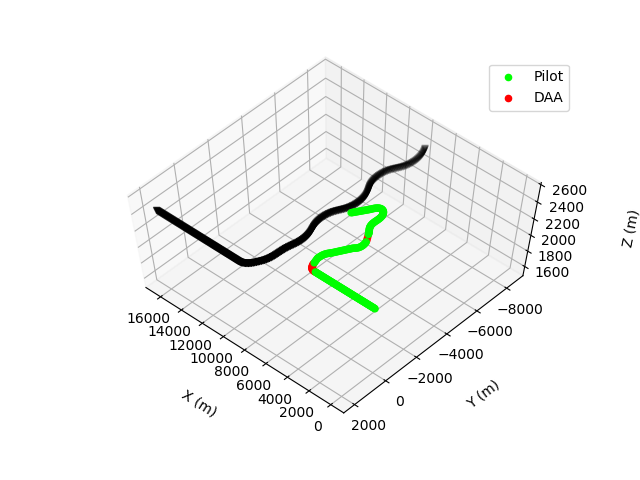}
}
\hfil
\subfloat(c){\includegraphics[width=2in]{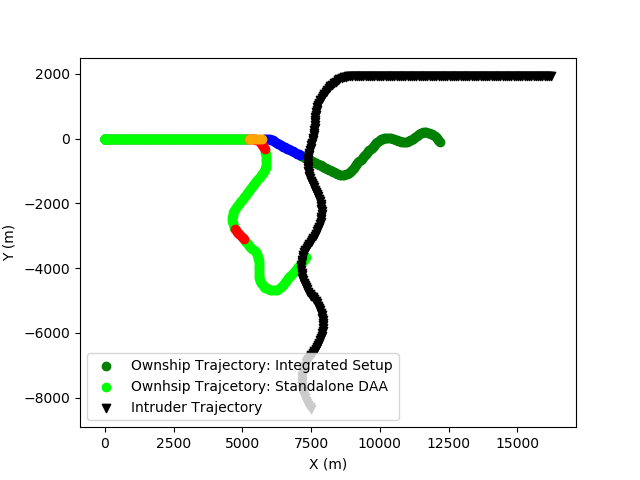}}
\hfil
\subfloat(d){\includegraphics[width=2in]{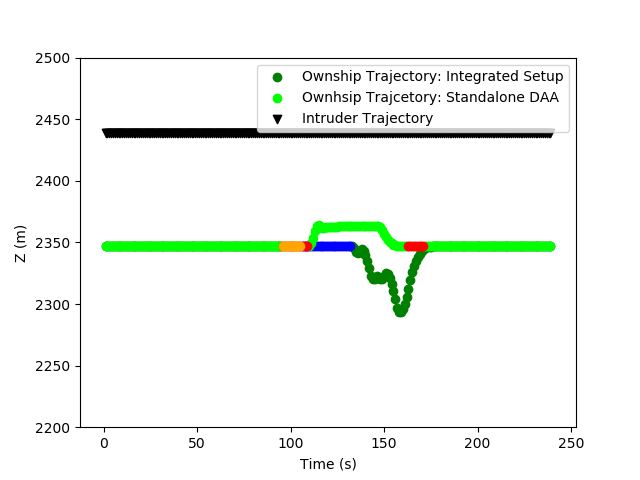}}

\centering
\caption{Trajectory analysis with and without the allocation agent in a head-on encounter, the black trajectory represents the intruder. (a) 3-D trajectories with the allocation agent, the orange dots represent the waiting status and the blue dots represent the blended command execution. (b) 3-D trajectories without the allocation agent, the red dots represent the onboard DAA execution. Note that the onboard DAA generates a different maneuver from the pilot's maneuver shown in (a).  (c) The horizontal view illustrates the differences in the trajectories.  (d) The vertical view illustrates the vertical  differences in the trajectories.  }
\label{fig:compare command track}
\end{figure*}

We inspect command reception and execution in each group and distinguish the changes between the groups. We would like to increase the reception of pilot commands before override taken by the onboard DAA and incorporate the pilot commands to enhance pilot involvement in the resolution process. The waiting action can potentially allow the UAS to receive more pilot commands before a resolution is made. To illustrate how integrating the agent helps pilot and decrease DAA override instances without jeopardizing safety, we provide a representative encounter in Figure \ref{fig:compare command track}.

From the Figure~\ref{fig:compare command track}-c and 8-d, the onboard DAA system without the allocation agent takes over the control and ends up with a trajectory more deviated than the trajectory produced by the allocation agent. Although the pilot regains control after the DAA takes over, the pilot maneuver changes depending on the situation. With the integrated setup, the UAS waits for the pilot command. When the delayed pilot command is received, the agent invokes the blending algorithm and executes the blended command. As the pilot's intention is incorporated into the command, upon the pilot's regaining control, the recovery maneuver is more refined and more on track. With the standalone DAA setup, the DAA takes over the control twice over the encounter. Although in both setups the UAS regains the altitude, the recovery trajectory is not achieved in the DAA setup whereas in the integrated setup the UAS is on track to the original path. The integrated agent improves both the resolution trajectory and the recovery trajectory in this encounter.

The numbers  of wait/no wait decisions for each configuration is illustrated in Table \ref{tab:waiting status}. The first two groups $B-1$ and $B-2$ do not have the allocation agent available in the system. Therefore, during active encounters if a pilot command is unavailable, the system immediately allocates the control to the onboard DAA. With the integrated setup, we have three groups of experiments: the $IC$ group refers to the encounters with constant wait time and the $ID$ groups refer to the encounters that utilize the wait maps generated from the MDP. The experiment groups in the integrated setups demonstrate how  waiting improves the reception of pilot commands and the utility of using a wait map instead of a constant wait time.   

\begin{figure*}[!t]
\centering
\subfloat(a){\includegraphics[width=1.8in]{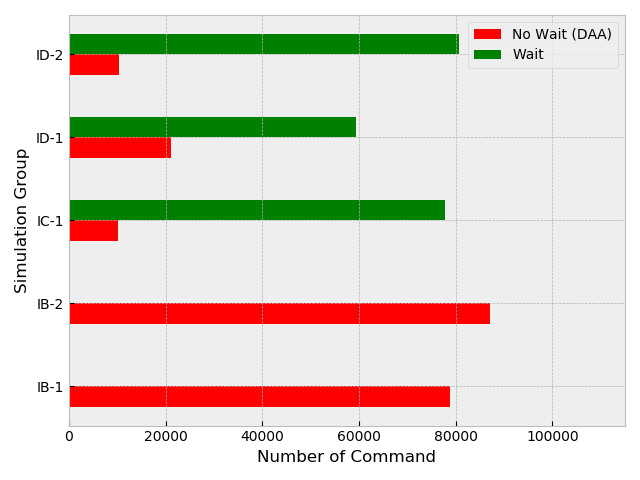}
}
\hfil
\subfloat(b){\includegraphics[width=1.8in]{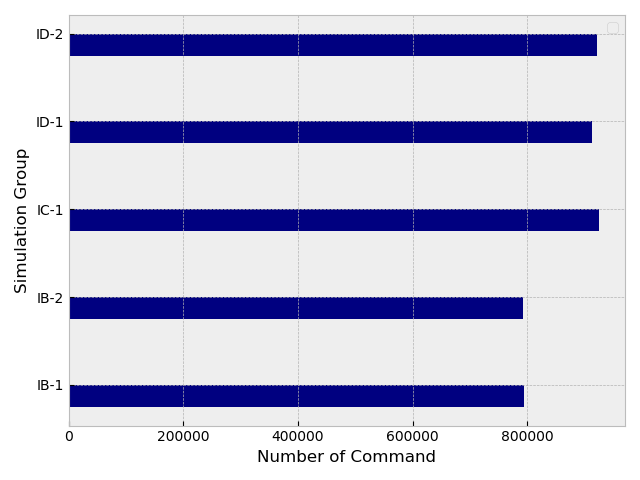}}
\hfil
\subfloat(c){\includegraphics[width=1.8in]{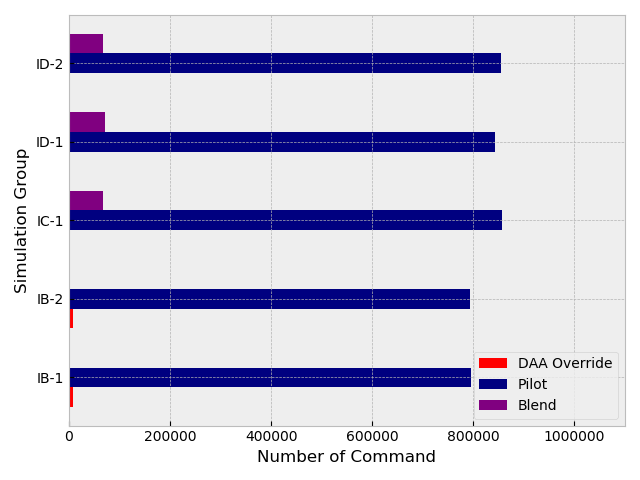}}
\centering
\caption{(a) Illustration of wait and no wait (allocate control to DAA) across different groups. The red bar represents onboard DAA commands. Data shows significant reduction in allocation to onboard DAA. (b) Illustration of pilot command reception.  (c) Bar graph showing command execution status. Integrated setups have reduced DAA override to zero. }
\label{fig:wait and pilot_receive and all}
\end{figure*}

On average, the reduction in the DAA allocation between the baseline and the integrated setup is $83.29\%$. We consider the $ID-2$ group as the comparison standard, as it may represent a more realistic environment, where the system does not have any knowledge of the latency duration. In Figure \ref{fig:wait and pilot_receive and all}-a, the bar plot illustrates the decrease in the DAA allocation, indicating  that the $ID-2$  group has significantly lowered the DAA allocation instances than $B-2$. This significant decrease in the DAA allocation instances results in an increases in the reception of pilot commands before the resolution is initiated by the onboard 
DAA. 

 \begin{table}[h!]

   \caption{Waiting Status Within Group}
    \label{tab:waiting status}
    \centering
    \begin{tabular}{c c c c }
    
         Groups  & Wait & No Wait (Allocate to DAA)  & Percent Reduction  \\
         \hline \hline
         B-1  & 0 & 78842 & - \\
         B-2 & 0 & 87153 & -\\
        IC-1 & 77732 & 10229 & 88.26\% \\
         ID-1 & 59295 & 21078 & 75.81\% \\
         ID-2 & 80650 & 10278 & 88.20\% \\

    \end{tabular}
   \footnotesize{*Percent Reduction with respect to $B-2$.} 
\end{table}

A summary of command generation and reception is presented in Table \ref{tab:pilot receive}. With the integrated setup, the pilot command reception is increased by \textbf{$15.81\%$} on average compared to the baseline setup. Comparing the time varying latency groups $B-2$ and $ID-2$, we see that the pilot command reception is increased by \textbf{$16.18\%$}, which includes the commands received before the override resolution.

\begin{table}[h!]

   \caption{Pilot Command Reception}
    \label{tab:pilot receive}
    \centering
    \begin{tabular}{c c c}
         Groups &  Number of Pilot Command Received & Percent Increment \\
         \hline \hline 
         B-1 & 795163 & - \\
         B-2 & 793529 & -\\
         IC-1 & 924574 & 16.19\% \\
         ID-1 & 913403 & 15.10\% \\
         ID-2 & 921993 & 16.18\%
    \end{tabular}
    \footnotesize{*Percent Increment with respect to $B-2$.} 
\end{table}

The bar graph in Figure \ref{fig:wait and pilot_receive and all}-b visually illustrates the change in command reception compared to our standard group $ID-2$. The pilot command reception statistics for the integrated setup in different groups are close. In the group $IC-1$, the simulator has prior knowledge of the latency that results in a higher number of pilot command reception. Hence, $IC-1$ has a slightly higher number of reception than our standard $ID-2$.

The increase in the pilot command reception also results in an increase in pilot involvement in resolution. As the system receives the delayed pilot command, the execution of the command depends on the safety metric that evaluates the feasibility and risk associated with the intended maneuver from the pilot. A positive safety metric indicates a safe maneuver. In the event where the received pilot command is unsafe, the augmented blending algorithm integrates the pilot intended maneuver types to generate DAA maneuvers. Table \ref{tab:all command} summarizes the command execution statistics. In the table, a DAA override is an instance where the onboard DAA takes the authority even though a pilot command is available. Figure \ref{fig:wait and pilot_receive and all}-c visually represents the data in different groups. 
  
\begin{table}[h!]

   \caption{Command Execution Statistics}
    \label{tab:all command}
    \centering
    \begin{tabular}{c c c c }
    
         Groups  & DAA override & Pilot & Blending    \\
         \hline \hline 
         B-1  & 8344 & 795163 & 0 \\
         B-2 & 7564 & 793529 & 0 \\
         IC-1& 0& 857648 & 66926 \\
         ID-1 & 0 & 842914 & 70489\\
         ID-2 & 0 & 853974 & 68019\\

    \end{tabular}
 
\end{table}
The proposed blending algorithm allows the onboard DAA to choose and execute a maneuver that closely matches the pilot preferred maneuver. Executing a maneuver preferred by the pilot is also expected to achieve a better recovery trajectory whenever the pilot regains the full control. We further analyze whether it is possible to find a matched maneuver. As the system waits, the encounter may go in a  closer proximity than it would go without waiting. On average, 84\% of blended commands were able to find and match the pilot's exact maneuver type and perform a resolution maneuver in the plane and direction that the pilot had chosen. The remaining $16\%$ of the commands execute maneuvers in the pilot intended plane. In all the scenarios, the pilot preferences in terms of maneuver are augmented into the resolution maneuvers.

\subsection{Safety Analysis}
\begin{figure*}[!t]
\centering
\subfloat(a){\includegraphics[width=1.8in]{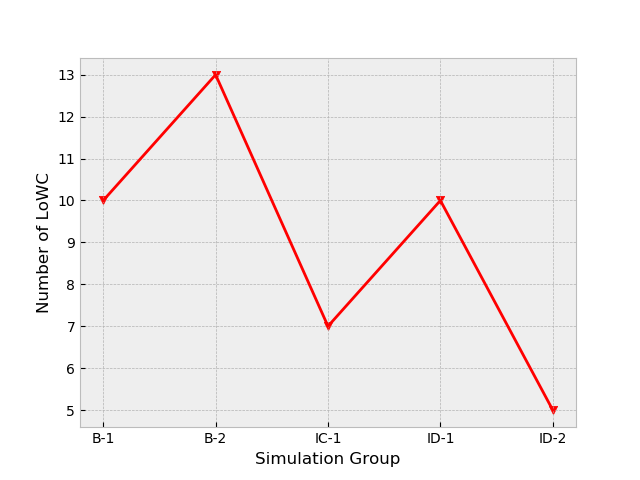}
}
\hfil
\subfloat(b){\includegraphics[width=1.8in]{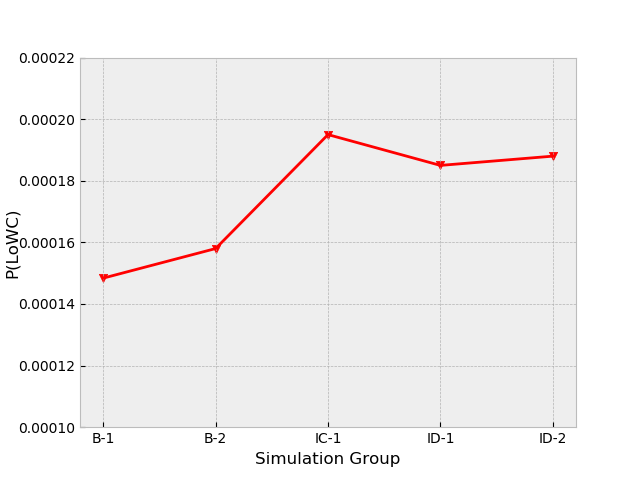}
}
\hfil
\subfloat(c){\includegraphics[width=1.65in]{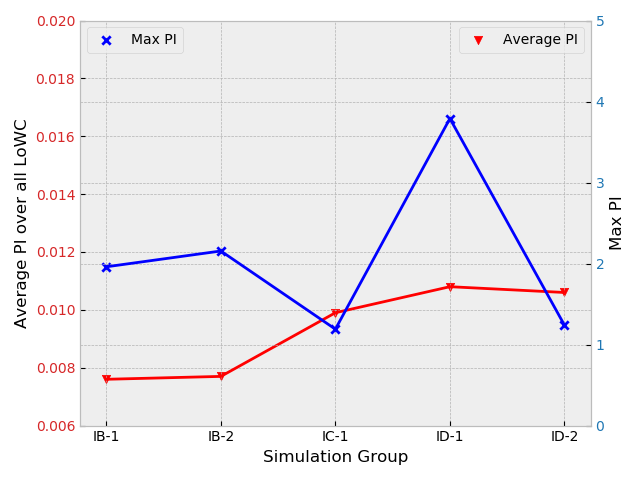}
}
\centering
\caption{(a) The number of LoWC encounters within 10,000 encounters, $ID-2$ group has least number of LoWC. (b) Probability of LoWC in sample trajectories in different groups. (c) Average PI in simulation groups and max PI in each group. }
\label{fig:lowc and ploss}
\end{figure*}

The number of LoWC occurrences in each group is illustrated in Figure \ref{fig:lowc and ploss}-a. The simulations in group $B-2$ exhibits the maximum number of LoWC occurrences, where $ID-2$ has the minimal number of LoWC occurrences. Note that we consider $ID-2$ as our standard. There are particular scenarios that suffer LoWC in multiple groups. On average, 98\% of the encounters that suffer LoWC belong to turning geometry where a high speed intruder converges towards the ownship. 

\begin{table}[h!]

   \caption{Likelihood definition~\cite{faahandbook}}
    \label{tab:faa}
   
    \centering
  \tiny
    \begin{tabular}{c c c }
     
         Frequency  & Qualitative & Quantitative P(instance)/flight hour \\
         \hline \hline 
         Frequent, A  & Expected to occur routinely &  within $1\times 10^{-3}$ to $1.00$ \\
         Probable, B  & Expected to occur often & 
within $1\times 10^{-3}$ to~$1\times10^{-5}$  \\
Remote, C & Expected to occur infrequently & within $1\times10^{-5}$ to~$1\times10^{-7}$ \\
Extremely Remote, D & Expected to occur rarely & within $1\times10^{-7}$ to~$1\times10^{-9}$ \\
Extremely Improbable, E & unlikely but it is not impossible & less than~$1\times10^{-9}$
\end{tabular}
 
\end{table}
 
 With a total of $666.67$ simulation hours (each of the $10,000$ encounters lasts $T=240$ seconds), we calculate the probability of LoWC occurrence per flight hour. The probability of LoWC is calculated using Monte Carlo approximation as
 \begin{equation}
     P(LoWC/flight~hr)= \frac{P(LoWC)}{Total~flight~hours~in~ simulation}.
 \end{equation}
   In Table~\ref{tab:faa}, we recall from the FAA System Safety Handbook~\cite{faahandbook} how quantitative values of $P(LoWC/flight~hr)$ can be correlated to qualitative measures using FAA safety measure definitions. The qualitative measure is summarized in Table \ref{tab:per flight los} and Figure \ref{fig:lowc and ploss}- b visually represents the probability of LoWC calculated from the simulation datasets.

\begin{figure*}[!t]
\centering
\subfloat(a){\includegraphics[width=1.8in]{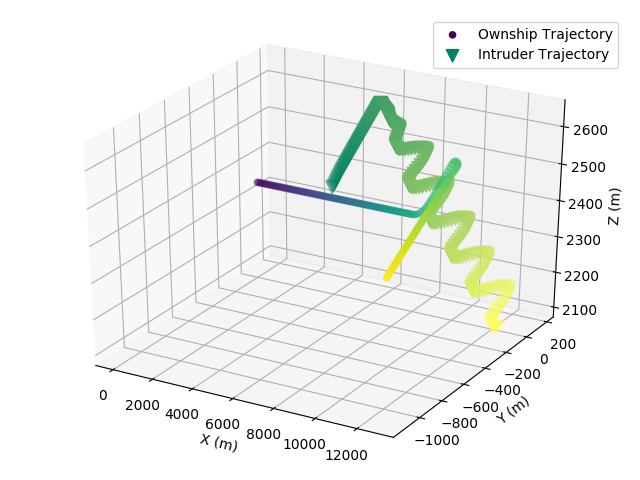}
}
\hfil
\subfloat(b){\includegraphics[width=1.8in]{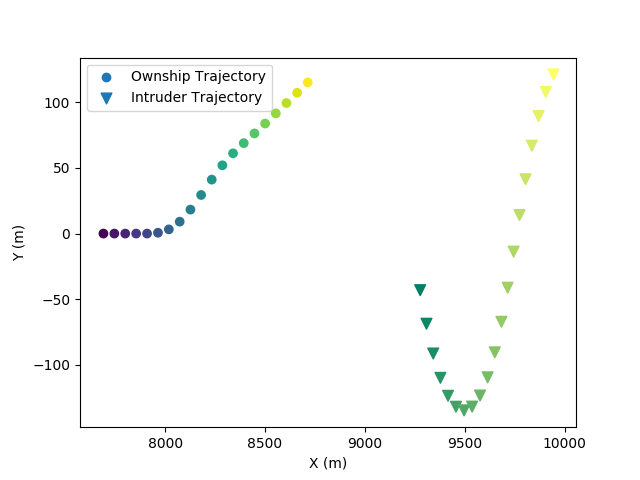}
}
\hfil
\subfloat(c){\includegraphics[width=1.8in]{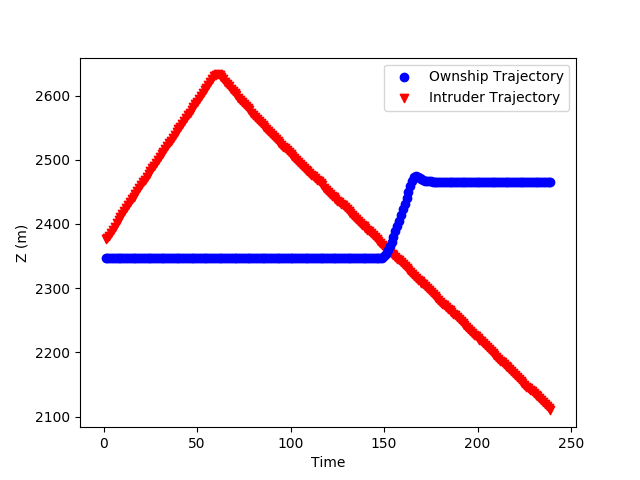}}

\centering
\caption{The flight geometry corresponding to the maximum PI.  (a) 3-D trajectory: color variation corresponds to time where darker shade indicates starting of the encounter. (b) minimum closure during instantaneous LoWC (horizontal view). (c) Vertical view.}
\label{fig:PI geometry}
\end{figure*}

\begin{table}[h!]

   \caption{Risk Measure per flight hour}
    \label{tab:per flight los}
    \centering
    \begin{tabular}{c c c }
    
         Groups  & P(LoWC)/flight hr & Qualitative Likelihood    \\
         \hline \hline 
         B-1  & $2.22\times10^{-7}$ &  Remote \\
         B-2 & $2.36\times10^{-7} $&  Remote \\
         IC-1&  $2.92\times10^{-7}$ &  Remote \\
         ID-1 &  $2.77\times10^{-7}$  &  Remote\\
         ID-2 & $2.81\times10^{-7}$   &  Remote\\

    \end{tabular}
 
\end{table}
The qualitative significance of occurrence according to~\cite{faahandbook} is that the incidents are expected to occur infrequently and considered remote. With the integrated setup, the system is able to keep the safety margin similar to the DAA only system (without waiting). Although the system results in a slight increase in the probability of the $LoWC$ occurrences, the system is able to maintain the safety margin in the same category as the standalone DAA system. This indicates the efficiency of the allocation agent and supports that waiting does not jeopardize safety and degrade risk measure value any more than the baseline system.   As all the experiment groups have suffered from LoWC occurrences, we have quantified the severity of each LoWC using the Penetration integral (PI). It measures the severity of LoWC with a positive number. The lower the value, the more instantaneous the LoWC. An illustration of the average PI over each simulation group and the maximum PI value in each simulation is provided in Figure \ref{fig:lowc and ploss}-c. The average PI for each simulation group is less than 1 with the maximum value of 3.79 in the $ID-1$ group. We further investigate the encounter with the maximum PI and notice that the geometry is very unique, where the intruder has a upside down $V$ trajectory as presented in Figure~\ref{fig:PI geometry}. However, this LoWC can also be considered only instantaneous and benign  as the value is significantly lower than 10~\cite{do2017365}.

We also examine whether the wait actions with the integrated setup induce more trajectory deviation. The mean trajectory deviations in the horizontal and vertical directions are provided in Table \ref{tab:trajectory deviation}. The results show that the mean trajectory deviation is almost the same for every setup. In the integrated setup, the UAS waits for the pilot command to arrive, which may induce more aggressive maneuvers, causing the UAS to deviate more from the original trajectory. However, we also observe that some encounters exhibit much higher deviations with the baseline setup.
 
\begin{table}[h!]

   \caption{Trajectory Deviation}
    \label{tab:trajectory deviation}
    \centering
    \begin{tabular}{c c c}
    
         Groups  & Horizontal deviation, m & Vertical deviation,m    \\
         \hline \hline 
         B-1  & 1690 & 184 \\
         B-2 & 1693 & 183 \\
         IC-1&  1690 & 183 \\
         ID-1 &  1693 & 182\\
         ID-2 & 1689 & 182 \\

    \end{tabular}
 
\end{table}

\section{Conclusions}\label{sec: conclusion}
 In this paper, we implement an allocation agent in the mixed initiative DAA-pilot platform that dynamically authorizes control between pilot and onboard DAA and blends the received pilot commands in the presence of communication latency. One of the main challenges in the presence of communication latency is the lack of situational awareness and equivalent level of safety without a pilot onboard. The wait map developed from the MDP can potentially allow safe waiting and incorporate more pilot commands in the DAA resolution. To further accommodate pilot commands, a blending algorithm augmenting pilot preference into DAA generated maneuvers is proposed. The integrated system positively improves pilot involvement into the resolutions and decreases override of pilot commands. The results of fast-time Monte Carlo simulations with random latency reflects the effectiveness of the wait maps and the blending algorithm. The integrated system maintains a similar safe margin as the standalone DAA system used in the simulation.

\bibliographystyle{elsarticle-num}
\bibliography{ref_journal.bib}





\end{document}